\documentclass[aps,twocolumn,pra,showpacs,tightenlines]{revtex4}
\usepackage{epsfig,graphicx,times}
\usepackage{amstext}
\usepackage{amsmath}
\usepackage{amssymb}
\usepackage{graphicx}
\usepackage{latexsym}
\usepackage{bm}
\usepackage[colorlinks,citecolor=blue,linkcolor=blue,hyperindex,dvipdfm]{hyperref}
\usepackage{subfigure}

\begin{document}

\title{The initial-state dependence of quantum speed limit}
\author{Shao-xiong Wu}
\author{Yang Zhang}
\author{ Chang-shui Yu}
\email{quaninformation@sina.com}
\author{He-shan Song}
\affiliation{School of Physics and Optoelectronic Technology, Dalian University of
Technology, Dalian 116024, China }
\date{\today }

\begin{abstract}
The generic bound of quantum speed limit time (the minimal evolution time)
for a qubit system interacting with structural environment is investigated. We define a new bound for the quantum speed limit. It is shown that the non-Markovianity and the population of the excited state can fail to signal the quantum evolution acceleration, but the initial-state dependence is an important factor. In particular, we find that different quantum speed limits could produce contradictory predictions on the quantum evolution acceleration.
\end{abstract}

\maketitle

\section{Introduction}

The maximal dynamical speed of quantum system is a fundamental concept in
many areas of quantum physics, such as quantum communication \cite%
{com-prl,com-nature}, quantum metrology \cite{metrology-prl,metrology-np},
optimal control \cite{control}, etc. The quantum speed limit (QSL),
defined as the minimal evolution time between two states was first
introduced by Anandan-Aharonov using the Fubini-Study metric \cite{Aharonov}%
. Later, the unified QSL is given for closed systems subject to the unitary evolution, by the
combining the conclusion of the variance of the energy, i.e., Mandelstam-Tamm
bound (MT-QSL) \cite{MT,Fleming,Aharonov,Vaidman,jpa} and the average of the
energy, i.e., Margolus-Levitin bound (ML-QSL) \cite{ML,VG,JK}, which is rigorously described by $%
t_{QSL}=\max\{\pi\hbar/(2\Delta E),\pi\hbar/(2E)\}$. The QSL bound for the
driven system \cite{LutzJPA,epl,drivingqsl,DeffnerJPB}, and for the mixed quantum state evolution\cite{Andersson1,Andersson2,HengFan} were also investigated. In reality, the system interacts with the its environment inevitably, so the theories of open quantum systems are usually employed \cite{Breuer}. Recently, the QSL for open system (nonunitary evolution) has attracted intensive interests, such as characterizing QSL for open system using quantum Fisher information \cite%
{qslfisher}, relative purity \cite{Plenio}, Bures angle \cite{LutzPRL}, and
so on \cite{xuzhenyucpb,yindu,unifiedbound,xuzhenyu,guohong,HengFan}. The works \cite{Plenio,qslfisher} also apply to non-Markovian dynamics. In
particular, in Ref. \cite{LutzPRL}, the authors extended the ML-QSL from
closed system to open system using the technology of the von Neuamnn trace
inequality \cite{vonneumann,vonneumann1} and operator norm \cite%
{Horn}, and showed that the non-Markovian effects \cite
{nonMarkovian,nonMarkovian1,SCHou,luoshunlong,nonMarkovian2008} can speed up
the quantum evolution and the ML-QSL is tight for the open system.  In Ref.
\cite{xuzhenyu}, the authors showed that the mechanism of the acceleration in open systems is determined not only by the non-Markovianity but also by the population of excited states during the quantum evolution.

In this paper, we find that the initial-state dependence is also a key factor to the acceleration of quantum evolution. In particular, the predications for the acceleration of quantum evolution based on different QSLs could be quite inconsistent. Here we first define a tight bound for the quantum speed limit without using the von Neumann trace inequality (We denote it by NI-QSL). Then we use the ML-QSL and the NI-QSL to investigate the damped Jaynes-Cumming (J-C) model with a superposition state as the initial state and the dephasing model. It is found that neither the non-Markovianity nor the population of the initial excited state is competent for the acceleration of quantum evolution, but the different initial states bring great effects. In addition, the ML-QSL and the NI-QSL sometimes produce contradictory predictions on the evolution acceleration, even though they are consistent in some particular cases. This implies that the QSL could need us further consideration and investigation.
This paper is organized as follows. In the Sec. \ref{sec2}, we give a brief introduction about the ML-QSL and define the NI-QSL. In Sec. \ref{sec3} and \ref{sec4}, we study the damped J-C model and the dephasing model, respectively. Some discussions
and conclusion are drawn at the end.

\section{The quantum speed limit}\label{sec2}
In Ref. \cite{LutzPRL}, the authors derived the ML-type quantum speed limit
bound for a pure state $\vert\psi _{0}\rangle $ based on the von Neumann trace
inequality and operator norm in the open system through geometric approach. The `distance'
 between the initial state $\rho_0=\vert\psi
_{0}\rangle\langle\psi_0\vert $ and its target state $\rho_t$
is measured by Bures angle, which reads
\begin{equation}
\mathcal{L}(\rho _{0},\rho _{t})=\arccos (\sqrt{\left\langle \psi
_{0}\right\vert \rho _{t}\left\vert \psi _{0}\right\rangle }).\label{Bruesangle}
\end{equation}%
The generalized time-dependent nonunitary equation can be expressed as $\dot{\rho}_{t}=L_{t}\left( \rho _{t}\right) $.
The time derivative on the Bures angle leads to the
relation
\begin{equation}
2\cos \left( \mathcal{L}\right) \sin \left( \mathcal{L}\right) \dot{\mathcal{%
L}}\leq \left\vert \left\langle \psi _{0}\right\vert \dot{\rho}%
_{t}\left\vert \psi _{0}\right\rangle \right\vert =\left\vert \text{tr}%
\{L_{t}\left( \rho _{t}\right) \rho _{0}\}\right\vert .  \label{tuidao1}
\end{equation}%
Based on the von Neumann trace inequality, the right-hand side of Eq. (\ref{tuidao1})
can arrive at
\begin{equation}
\left\vert {\mathrm{tr}\{{{L_{t}}\left( {{\rho _{t}}}\right) {\rho _{0}}}\}}%
\right\vert \leq \sum\limits_{i=1}^{n}{{\sigma _{1,i}}}{\sigma _{2,i}}
\label{vonneumann}
\end{equation}%
with $\sigma _{1,i}$ and $\sigma _{2,i}$ being the singular values of matrix
$L_{t}\left( \rho _{t}\right) $ and $\rho _{0}$ in decreasing order.
Integrating Eq. (\ref{tuidao1}) over time, one can obtain the ML-QSL for the nonunitary
generator as
\begin{equation}
t_{QSL}^{ML}\geq \frac{\sin ^{2}[\mathcal{L}(\rho _{0},\rho _{t})]}{\Lambda
_{t}^{\text{op}}},  \label{qslml}
\end{equation}%
where $\Lambda _{t}^{\text{op}}=(1/t)\int_{0}^{t}d\tau \left\Vert L_{t}(\rho
_{\tau })\right\Vert _{\text{op}}$ with $\left\Vert \cdot \right\Vert _{%
\text{op}}$ being the operator norm (the maximum singular value) of the matrix.

However, if the von Neumann trace inequality Eq. (\ref{vonneumann}) and operator norm are not used, we can directly obtain, through integrating Eq. (\ref{tuidao1}) over time,
\begin{equation}
t_{QSL}^{NI}\geq \frac{\sin ^{2}[\mathcal{L}(\rho _{0},\rho _{t})]}{%
(1/t)\int_{0}^{t}d\tau \left\vert \left\langle \psi _{0}\right\vert
L_{t}(\rho _{\tau })\left\vert \psi _{0}\right\rangle \right\vert },
\label{qslno}
\end{equation}
 which we call the NI-QSL to distinguish from the ML-QSL. The prominent feature of Eq. (\ref{qslno}) is that it provides a tighter bound than ML-QSL given by Eq. (\ref{qslml}).
Our main conclusion will be obtained by comparing the two types of QSL for the dynamics of open quantum systems.

\section{The quantum speed limit time for the damped Jaynes-Cumming model}\label{sec3}

The first model we will consider includes a single qubit interacting with vacuum reservoir \cite{Breuer}, which is
also called the damped J-C model. The whole Hamiltonian of the system and
reservoir is $H=\omega _{0}\sigma _{z}+\sum_{k}\omega _{k}b_{k}^{\dagger
}b_{k}+\sum_{k}(g_{k}\sigma _{+}b_{k}+g_{k}^{\ast }\sigma _{-}b_{k}^{\dagger
})$. The  dynamics of the system can be described by%
\begin{equation}
L_{t}(\rho _{t})=\gamma _{t}(\sigma _{-}\rho \sigma _{+}-\frac{1}{2}\sigma
_{+}\rho \sigma _{-}-\frac{1}{2}\sigma _{-}\rho \sigma _{+}).  \label{lt}
\end{equation}%
The structure of environment is assumed as the Lorentzian form:
\begin{equation}
J\left( \omega \right) =\sum_{k}\left\vert g_{k}\right\vert ^{2}\delta
(\omega _{0}-\omega _{k})=\frac{\gamma _{0}}{2\pi }\frac{\lambda ^{2}}{%
(\omega _{0}-\omega )^{2}+\lambda ^{2}},
\end{equation}%
where $\lambda $ is the spectral width of the reservoir and $\gamma _{0}$ is
the decay of the system. The Markovian and non-Markovian regimes
can be distinguished by the relation of parameters $\gamma $ and $\lambda $. In the Markovian regime, we have $\gamma _{0}<\lambda /2$ and
in the non-Markovian regime, we have $\gamma _{0}>\lambda /2$ \cite{nonMarkovian,nonMarkovian1,nonMarkovian2008,SCHou,luoshunlong}.

If the initial state of system is a superposition state
\begin{equation}
|\psi \rangle =\alpha e^{i\theta }|1\rangle +\sqrt{1-\alpha ^{2}}|0\rangle ,
\label{intialstate}
\end{equation}%
the final reduced state after the evolution will be \cite{Breuer}
\begin{equation}
\rho _{t}=\left(
\begin{array}{cc}
\alpha ^{2}|q(t)|^{2} & \alpha \sqrt{1-\alpha ^{2}}\text{e}^{i\theta }q(t)
\\
\alpha \sqrt{1-\alpha ^{2}}\text{e}^{-i\theta }q^{\ast }(t) & 1-\alpha
^{2}|q(t)|^{2}%
\end{array}%
\right) ,
\end{equation}%
where $q(t)$ is determined by the integro-differential equation $\dot{q}%
(t)=\int_{0}^{t}d\tau f(t-\tau )q(\tau )$ with the correlation kernel
related to the spectral density of the reservoir as $f(t-\tau )=\int d\omega
J(\omega )e^{i(\omega _{0}-\omega )(t-\tau )}$. Using the Laplace transformation
and its inverse transformation, $q(t)$ can be given by
\begin{equation}
q(t)=e^{-\frac{\lambda t}{2}}\left[ \cosh \left( \frac{dt}{2}\right) +\frac{%
\lambda }{d}\sinh \left( \frac{dt}{2}\right) \right] ,
\end{equation}%
with $d=\sqrt{\lambda ^{2}-2\lambda \gamma _{0}}$. The time dependent decay
rate $\gamma _{t}$ in the Eq. (\ref{lt}) is given by $\gamma _{t}=-%
\mathrm{{Im}}\left(\frac{\dot{q}(t)}{q(t)}\right)$ .
After some calculations, we can obtain the ML-QSL for the initial state $|\psi \rangle $ as
\begin{equation}
t_{QSL}^{ML}=\frac{t\left\vert \alpha \right\vert (1-q(t))[1-(1-2\alpha
^{2})q(t)]}{\int_{0}^{t}\left\vert \sqrt{1-(1-4q(\tau )^{2})\alpha ^{2}}\dot{%
q}(\tau )\right\vert d\tau }.  \label{jcqslml}
\end{equation}%
\begin{figure}[b]
\centering
\includegraphics[width=0.75\columnwidth]{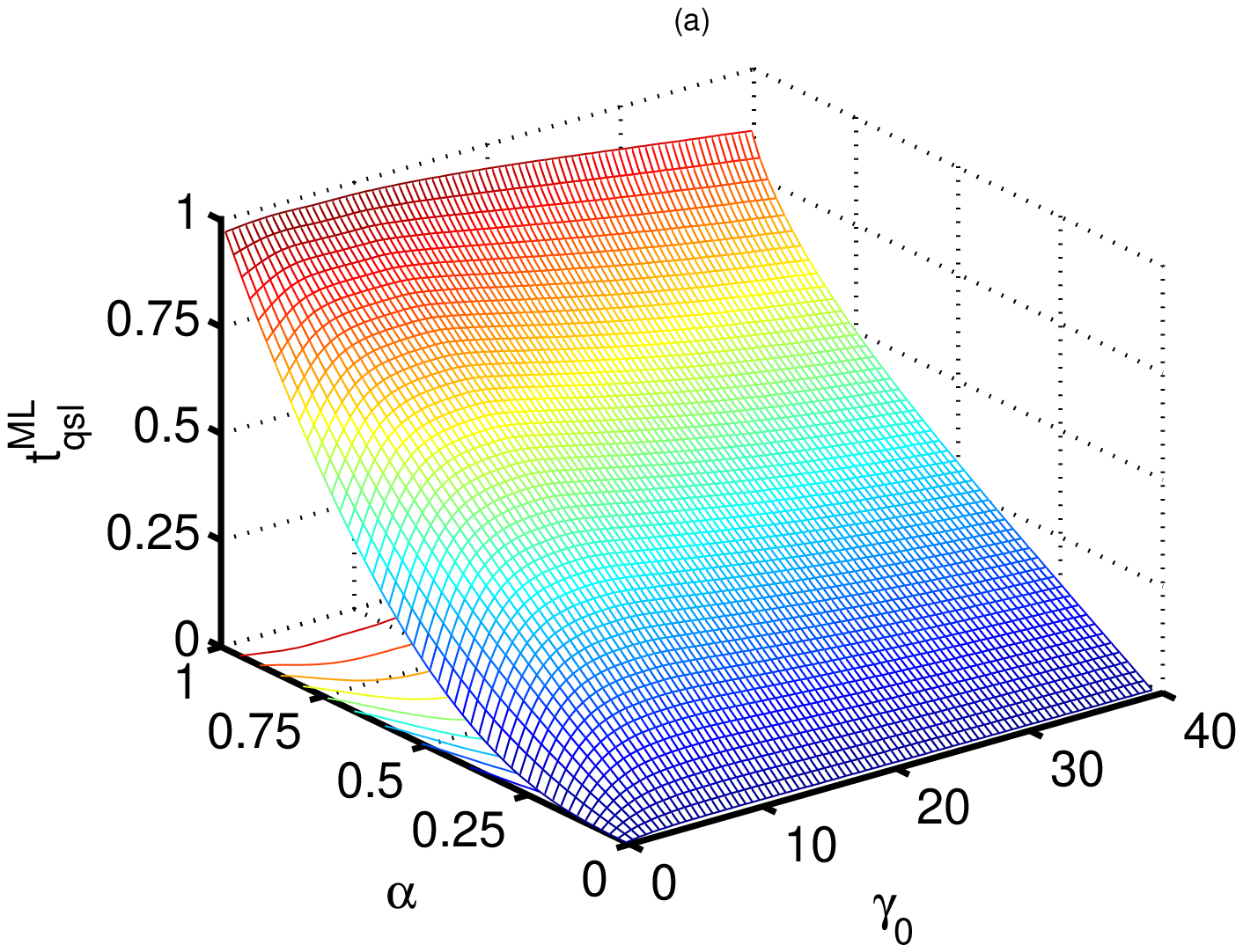} %
\includegraphics[width=0.75\columnwidth]{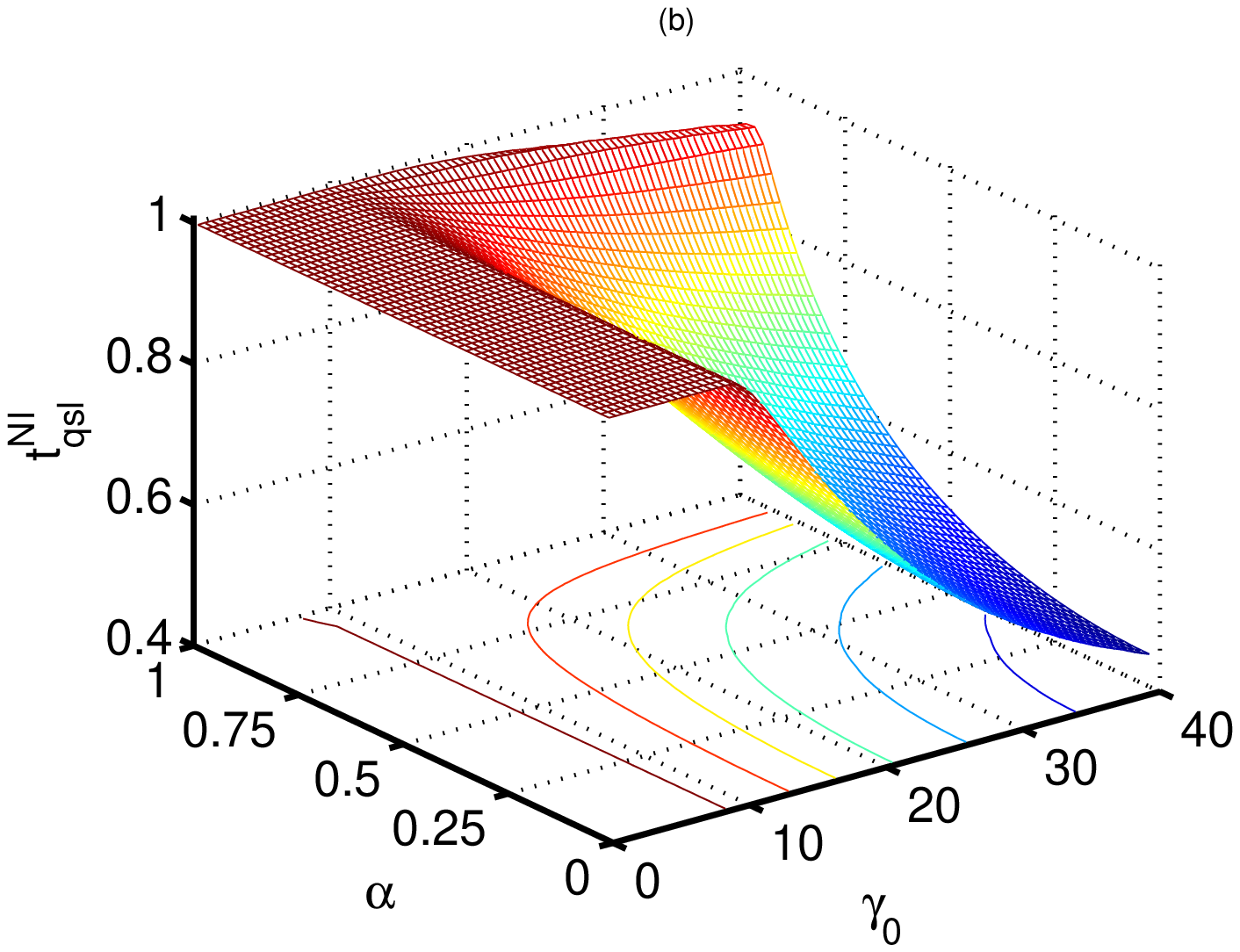}
\caption{(Color online). The quantum speed limit time for the initial state given by Eq. (%
\ref{intialstate}). Panel (a) is the ML-QSL $t_{QSL}^{ML}$, and Panel (b)
is the NI-QSL $t_{QSL}^{NI}$, where the parameter $\lambda=15$ and the actual evolution time is $t=1$.}
\label{jctu1}
\end{figure}
From Eq. (\ref{jcqslml}), one can find that the ML-QSL $t_{QSL}^{ML}$ is related to the $\alpha $, $q(t)$ and $\dot{q}(t)$. The variation of ML-QSL $t_{QSL}^{ML}$ with the parameter $\gamma _{0}$ and $\alpha $ is plotted in Panel (a) of Fig. \ref{jctu1}. In Fig. \ref{jctu1}, the actual evolution time is chosen as $t=1$ and  $\lambda=15$. $%
t_{QSL}^{ML}$ is symmetry about the population parameter $\alpha =0$, so we
just study the part $\alpha >0$. It is obviously shown that for $\gamma_0<\lambda/2$, the quantum speed limit time is below the actual evolution time $1$. Namely, the quantum evolution displays the acceleration in the Markovian regime. This is quite different from that the acceleration only appears in the non-Markovian regime due to non-Markovianity which was discussed $\alpha=1$ in Ref. \cite{LutzPRL}. One could imagine that the non-Markovianity is not the only reason.  The population of the excited state is also a factor for the such an acceleration as that given in Ref. \cite{xuzhenyu}, which is determined by
\begin{equation}
t_{QSL}=\frac{t}{2\frac{N\left( L_t(\rho_t)\right) }{1-\left\vert
q(t)\right\vert ^{2}}+1},  \label{zhushiqun}
\end{equation}%
where $N\left( \cdot\right) $ is the degree of non-Markovianity for
dynamics defined as the total backflow of information  \cite{nonMarkovian} and $\left\vert q(t)\right\vert ^{2}$ is the population of  the excited state, one should notice that in the Markovian regime, $N\left( \cdot\right) =0$, it will directly eliminate the role of the population $\left\vert q(t)\right\vert ^{2}$ in Eq. (\ref{zhushiqun}). Thus one possible reason for the acceleration subject to the ML-QSL is the dependence of the initial state. The details on the non-Markovianity is given by the Appendix A. 

\begin{figure}[t]
\centering
\includegraphics[width=0.75\columnwidth]{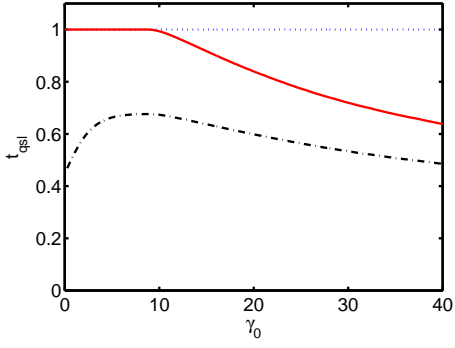}
\caption{(Color online). The quantum speed limit time for the initial state $\left\vert\psi\right \rangle =(\left\vert 1\right\rangle +\left\vert 0\right\rangle )/\sqrt{2}$. Here $\lambda=15$. The dashed line is the actual evolution time $t=1$, the dash-dotted line stands for the ML-QSL $t_{QSL}^{ML}$ and the solid line represents the NI-QSL $t_{QSL}^{NI}$. }
\label{jctu2}
\end{figure}

If we use the NI-QSL to characterize the quantum speed limit,  we have to calculate the NI-QSL for the initial state $\vert\psi\rangle $ as
\begin{equation}
t_{QSL}^{NI}=\frac{t(1-q(t))[1-(1-2\alpha ^{2})q(t)]}{\int_{0}^{t}\left\vert
2[1-q(\tau )-(1-2q(\tau )^{2})\alpha ^{2}]\dot{q}(\tau )\right\vert d\tau }.
\label{jcqslno}
\end{equation}
The variation of ML-QSL $t_{QSL}^{ML}$ with the parameter $\gamma _{0}$ and $\alpha $ is plotted in Panel (b) of Fig. \ref{jctu1}. One can find that the quantum speed limit time is fixed to the actual
evolution time in the Markovian regime, while the evolution acceleration is demonstrated  in the non-Markovian regime, i.e., $\gamma _{0}>\lambda /2$. In order to give a more
intuitive illustration, the QSLs $t_{QSL}^{ML}$ and $t_{QSL}^{NO}$ for
the system with initial state $\vert\psi \rangle =\frac{1}{\sqrt{2}} (|1\rangle +|0\rangle )$ are plotted in Fig. \ref{jctu2}. In Fig. \ref{jctu2}, the parameters are chosen the same as Fig. \ref{jctu1}. The initial-state dependence can also been found from Fig. \ref{jctu1} by different $\alpha$.

Comparing the quantum speed limit $t_{QSL}^{ML}$ and $t_{QSL}^{NI}$, one can find that  $t_{QSL}^{NI}=t_{QSL}^{ML}$ for the initial state $\left\vert 1\right\rangle $ (i.e., $\alpha=1$). However, if $\alpha\neq 1$, $t_{QSL}^{ML}$ and $t_{QSL}^{NI}$ usually demonstrates different behaviors which has been analyzed previously. Thus the ML-QSL and the NI-QSL produce contradictory predictions on the evolution acceleration.

\section{Quantum speed limit for the dephasing model}\label{sec4}

In the following, we will consider another exactly solvable model,
a two-level system coupling with a harmonic oscillator reservoir, it is
also called the dephasing model \cite{Breuer}. In the Schr\"{o}dinger picture, the total
Hamiltonian is taken to be $H=\frac{\omega _{0}}{2}\sigma
_{z}+\sum_{k}\omega _{k}b_{k}^{\dagger }b_{k}+\sum_{k}\sigma
_{z}(g_{k}b_{k}^{\dagger }+g_{k}^{\ast }b_{k})$. The evolution operator of
the system is $L_{t}(\rho _{t})=\frac{\gamma _{t}}{2}%
(\sigma _{z}\rho _{t}\sigma _{z}-\rho _{t})$. The initial state is
\begin{equation}
|\phi \rangle =\beta e^{i\theta }|1\rangle +\sqrt{1-\beta ^{2}}|0\rangle ,\label{spininitialstate}
\end{equation}
and the dynamics of the reduced system is expressed as \cite{Breuer}
\begin{equation}
\rho _{t}=\left(
\begin{array}{cc}
\beta ^{2} & \beta \sqrt{1-\beta ^{2}}e^{i\theta -\gamma (t)} \\
\beta \sqrt{1-\beta ^{2}}e^{-i\theta -\gamma (t)} & 1-\beta ^{2}%
\end{array}%
\right) .
\end{equation}
Taking the continuum limit of the bath mode and introducing the spectrum $J(\omega)$ of
the environment, we can find the dephasing factor $\gamma (t)$ given by
\begin{equation}
\gamma (t)=\int_{0}^{\infty }d\omega J(\omega )\coth (\frac{\omega }{2k_{B}T}%
)\frac{1-\cos {\omega t}}{\omega ^{2}}.
\end{equation}%
The spectrum of the environment is chosen as the Ohmic-like spectrum with
soft cutoff \cite{Weiss,Haikka}
\begin{equation}
J(\omega )=\eta \frac{\omega ^{s}}{\omega _{c}^{s-1}}\exp (-\omega /\omega
_{c}),
\end{equation}%
where $\omega _{c}$ is the cutoff frequency, $\eta $ is the
dimensionless coupling constant and the parameter $s>0$.
For simplicity, we will assume that the cutoff frequency $%
\omega _{c}$ is $1$.  $s$ determines the property of the environment
such as the sub-Ohmic reservoir for $s<1$, the Ohmic reservoir for $s=1$ and the super-Ohmic
reservoir for $s>1$. Under the condition $T=0$, $t>0$ and $s>0$, the
dephasing factor $\gamma (t)$ can be obtained \cite{chin} by
\begin{equation}
\gamma (t)=\eta \left[ 1-\frac{\cos [(s-1)\arctan (t)]\Gamma (s-1)}{%
(1+t^{2})^{(s-1)/2}}\right],
\end{equation}%
where $\Gamma (\cdot )$ is the Euler Gamma Function.
Thus the ML-QSL for the state in Eq. (\ref{spininitialstate}) can be given by
\begin{equation}
t_{QSL}^{ML}=\frac{2\beta \sqrt{1-\beta ^{2}}(1-e^{-\gamma (t)})t}{%
\int_{0}^{t}\left\vert e^{-\gamma (\tau )}\gamma ^{\prime }(\tau
)\right\vert d\tau }.  \label{spinqslml}
\end{equation}

\begin{figure}[b]
\centering
\includegraphics[width=0.75\columnwidth]{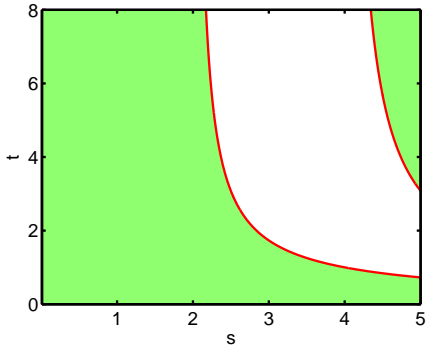}
\caption{(Color online). The illustration of the sign of the dephasing decay rate $\protect%
\gamma ^{\prime }(t)$. The green region stands for
the positive part of the $\protect\gamma ^{\prime }(t)$ which corresponds to the Markovian region, i.e., the
inequality Eq. (\ref{spinbudengshi}) $k\protect\pi <s\arctan (t)<(2k+1)\protect\pi $ satisfied, and the blank region corresponds to the negative part of $\protect\gamma %
^{\prime }(t)$ which corresponds to the non-Markovian region. If $t\rightarrow\infty$, the left critical edge is $s=2$.}\label{spintu1}
\end{figure}

From Eq. (\ref{spinqslml}), one can find that the ML-QSL $t_{QSL}^{ML}$ depends on the dephasing rate $\gamma'(t) $, which is the time derivative of the dephasing factor $\gamma(t)$ and determined by
  \begin{eqnarray}
  \gamma'(t)=\int d\omega J(\omega )\frac{\sin(\omega t)}{\omega}.
  \end{eqnarray}
Here, we have used the zero temperature condition.  In addition, one can see that, if $\gamma ^{\prime}(t)>0$, the integral in denominator of Eq. (\ref{spinqslml}) can be given analytically
as $1-e^{-\gamma (t)}$. It will obviously lead to that $t_{QSL}^{ML}=2\beta\sqrt{1-\beta^2}t$. In other words, the quantum speed limit time is fixed for the given initial state. It is irrelevant to the property of the environment.  However, in the region of $\gamma'(t)<0$, in usual the denominator of Eq. (\ref{spinqslml}) is hard to give explicitly. So it is important to tell the sign of $\gamma'(t)$ in order to give the ML-QSL. For the Ohmic-type spectral, the dephasing rate $\gamma'(t)$ can be given analytically as
\begin{equation}
\gamma ^{\prime }(t)=\eta (1+t^{2})^{-s/2}\Gamma (s)\sin (s\arctan (t)).\label{spinbudengshi}
\end{equation}%
One can find that $\gamma ^{\prime }(t)$ is always positive
if $\sin (s\arctan (t))>0$, i.e., $s$ and $t$ satisfy
\begin{equation}
k\pi <s\arctan (t)<(2k+1)\pi,
\end{equation}%
otherwise, $\gamma'(t)$ is negative. A vivid illustration of Eq. (\ref{spinbudengshi}) is provided in Fig. \ref{spintu1}. The green region corresponds to the positive $\gamma'(t)$, and the other black region  stands for the negative  $\gamma'(t)$.

\begin{figure}[t]
\centering
\includegraphics[width=0.75\columnwidth]{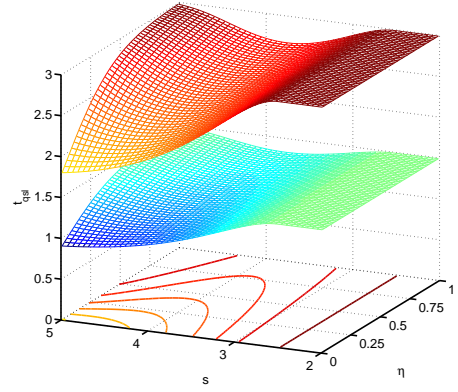}
\caption{(Color online). $t_{QSL}^{ML}$(including $t_{QSL}^{NI}$) versus with the Ohmic parameter $s$ and coupling
constant $\eta $.  The contour corresponds to the upper layer. At $s\simeq 2.5$, there is a straight contour, which signals no acceleration of quantum evolution for $s\lesssim 2.5$. The actual evolution time is $t=3$.}\label{spintu2}
\end{figure}

The ML-QSL $t_{QSL}^{ML}$ versus $s>2$ and $\eta$ is plotted in Fig. \ref{spintu2}, where the lower layer corresponds to $2\beta \sqrt{1-\beta ^{2}}=0.5$ and the upper layer corresponds to $\beta=\frac{1}{\sqrt{2}}$. It can be found that the ML-QSL for $\beta=\frac{1}{\sqrt{2}}$ serves the tightest  ML-QSL for all possible $\beta$. Here we choose the actual evolution time as $t=3$.
Since  $\arctan (t)\in [0,\pi /2)$, if $s\leq 2$,  $\gamma ^{\prime }(t)$ will always be positive, which can be seen from Fig. \ref{spintu1}. In this case, the ML-QSL  will be a constant and independent of  $s$ and $\eta $, for example, $t_{QSL}^{ML}=1.5$ for the lower layer and $t_{QSL}^{ML}=3$ is just the actual evolution time for the upper layer.  So the part for $s<2$ is not shown in Fig. \ref{spintu2}.  Compared with Fig. \ref{spintu1}, one can find that the ML-SQL keeps invariant until $s\simeq2.5$. For all $s$ the ML-SQL given by the lower layer shows the acceleration of the quantum evolution, but no acceleration is shown by the upper layer for $s\lesssim 2.5$ . In both layers, the population of the excited state is not changed due to the dephasing dynamics. Therefore, the acceleration with different degrees could be understood by the non-Markovianity. In Ref. \cite{Haikka}, it is shown that  for $s>2$ the dynamics for $t\rightarrow \infty$ enter the non-Markovian regime.  In fact, if $t$ is finite, one can easily find that the negative $\gamma'(t)$ will lead to the non-Markovian dynamics (see Appendix B for details). In other words, Fig. 3 just shows the bounds of the Markovian region and the non-Markovian region. However, the two layers in Fig. \ref{spintu2} corresponding to the different initial states imply the opposite properties of the acceleration especially for $2<s\lesssim 2.5$. Based on Fig. 3, the non-Markovianity could effectively signal the change of quantum evolution velocity for $\beta=\frac{1}{\sqrt{2}}$, but it fails for the lower layer. This shows the strong dependence of the initial state.
Similarly,  the NI-QSL can be given by
\begin{equation}
t_{QSL}^{NI}=\frac{(1-e^{-\gamma (t)})t}{\int_{0}^{t}\left\vert e^{-\gamma
(\tau )}\gamma ^{\prime }(\tau )\right\vert d\tau }.  \label{spinqslno}
\end{equation}
Comparing Eq. (\ref{spinqslml}) with (\ref{spinqslno}), one can immediately find that
the NI-QSL $t_{QSL}^{NI}$ does not depend on the population of the excited state. It is interesting that in this particular model, the NI-QSL does not depend on the initial state either. The ML-QSL $t_{QSL}^{ML}$ is connected with the NI-QSL $t_{QSL}^{NI}$ by a factor $2\beta\sqrt{1-\beta^2}$. Thus the NI-QSL versus $s$ and $\eta$ just consistent with the ML-QSL for $\beta=\frac{1}{\sqrt{2}}$, i.e., the upper layer in Fig. \ref{spintu2}. Since the opposite prediction on the evolution acceleration has been analyzed by the two layers, it is also implied that the NI-QSL and the ML-QSL demonstrate contradictory predictions.

\section{Discussions and Conclusion}

One may notice that the ML-QSL $t_{QSL}^{ML}$ is zero for some special
states, such as the ground state $\vert0\rangle$ for damped J-C model, the
excited state $\vert1\rangle$ or ground state $\vert0\rangle$ for
dephasing model. However, the `distance' (Bures angle) between the initial state and some other target state is also zero. So it is not difficult to understand the \textit{zero} evolution time. In addition, the  NO-QSL is obtained  without the von Neumann trace inequality and operator norm used, so it is obviously tighter than (at least, consistent with ) the previous bounds.

In summary, we have studied the ML-QSL and the NI-QSL of the damped J-C model and the dephasing model, and found that the importance of the initial-state dependence for the quantum evolution acceleration. To some extent, this is consistent with the derivation and conclusion the MT-QSL in the open system dynamics, which is evaluated with respect to the initial state \cite{Plenio}.  We also find that the predictions on the evolution acceleration based on different QSLs could produce contradictory conclusions. This implies that the QSL could deserve us further consideration. We hope that our work could deepen the understanding of the quantum speed limit on the dynamics of open systems.

\section*{Acknowledgement}

This work was supported by the National Natural Science Foundation of China,
under Grants No.11375036 and 11175033, and the Xinghai Scholar Cultivation Plan.

\section*{Appendix A: The non-Markovianity of damped J-C model}\label{appendixa}

Following Ref. \cite{nonMarkovian}, the non-Markovianity measure of the damped
J-C model we employed is based on the trace distance
$D(\Phi\rho _1,\Phi\rho_2)=\frac{1}{2}\mathrm{tr}\left\vert \Phi\rho _1-\Phi\rho_2\right\vert $. The change rate of the trace distance
is $\partial _{t}D(\Phi\rho _1,\Phi\rho_2)$. The positive change rate
stands for the flow of information from the environment back to the system.
The non-Markovianity of the quantum process $\Phi(t)$ can be given by%
\begin{equation}
N(\Phi)=\max_{\rho _{1,2}(0)}\int_{\partial _{t}D>0}dt\partial _{t}D(\Phi\rho _1,\Phi\rho_2).
\end{equation}
It is hard to obtain the optimal initial state pair ($\rho_1,\rho_2$) for a general
process. However, for the damped J-C model, it proves that the excited state $\left\vert 1\right\rangle $ and the ground state $\left\vert 0\right\rangle $ are the optimal state pair, so $\partial _{t}D(\Phi\rho _1,\Phi\rho_2)$
can be given by a simple expression $\partial _{t}D(\rho _{1}(t),\rho _{2}(t))=2\left\vert q(t)\right\vert\cdot \left\vert q(t)\right\vert^{\prime }$ with $\left\vert q(t)\right\vert^{\prime }$ being the time derivative of $\left\vert q(t)\right\vert$.
\begin{figure}[h]
  \centering
  \includegraphics[width=0.75\columnwidth]{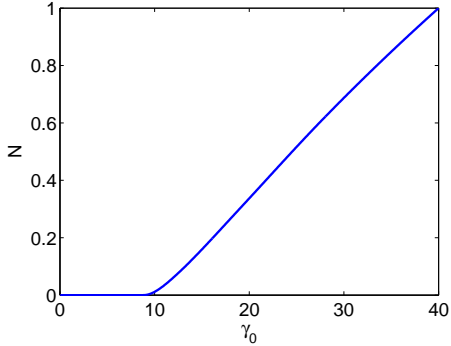}
  \caption{The non-Markovianity vs. $\gamma_0$. The parameter $\lambda=15$. The non-Markovianity measure is normalized to unity.}\label{ershenjc}
\end{figure}
Thus one can easily obtain the relation between the non-Markvianity and the parameter $\gamma _{0}$, which is plotted in Fig. \ref{ershenjc}. Comparing with Figs. \ref{jctu1} and \ref{jctu2}, one can find that the quantum evolution can be accelerated in the non-Markovian regime based on NI-QSL. However, based on the ML-QSL, the quantum evolution is accelerated not only in the non-Markovian regime but also in the Markovian regime.
\begin{figure}[t]
\centering
\includegraphics[width=0.75\columnwidth]{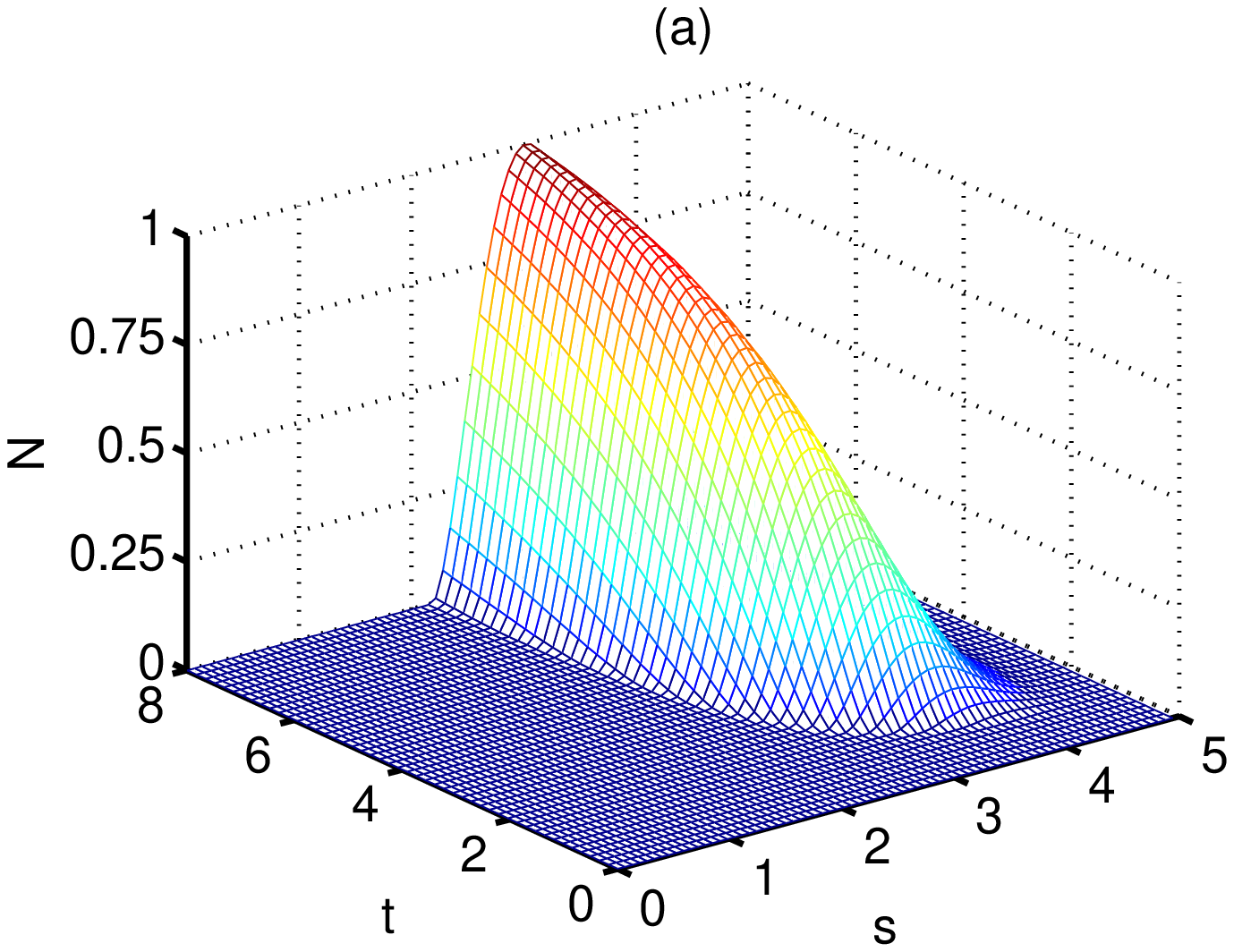}
\includegraphics[width=0.75\columnwidth]{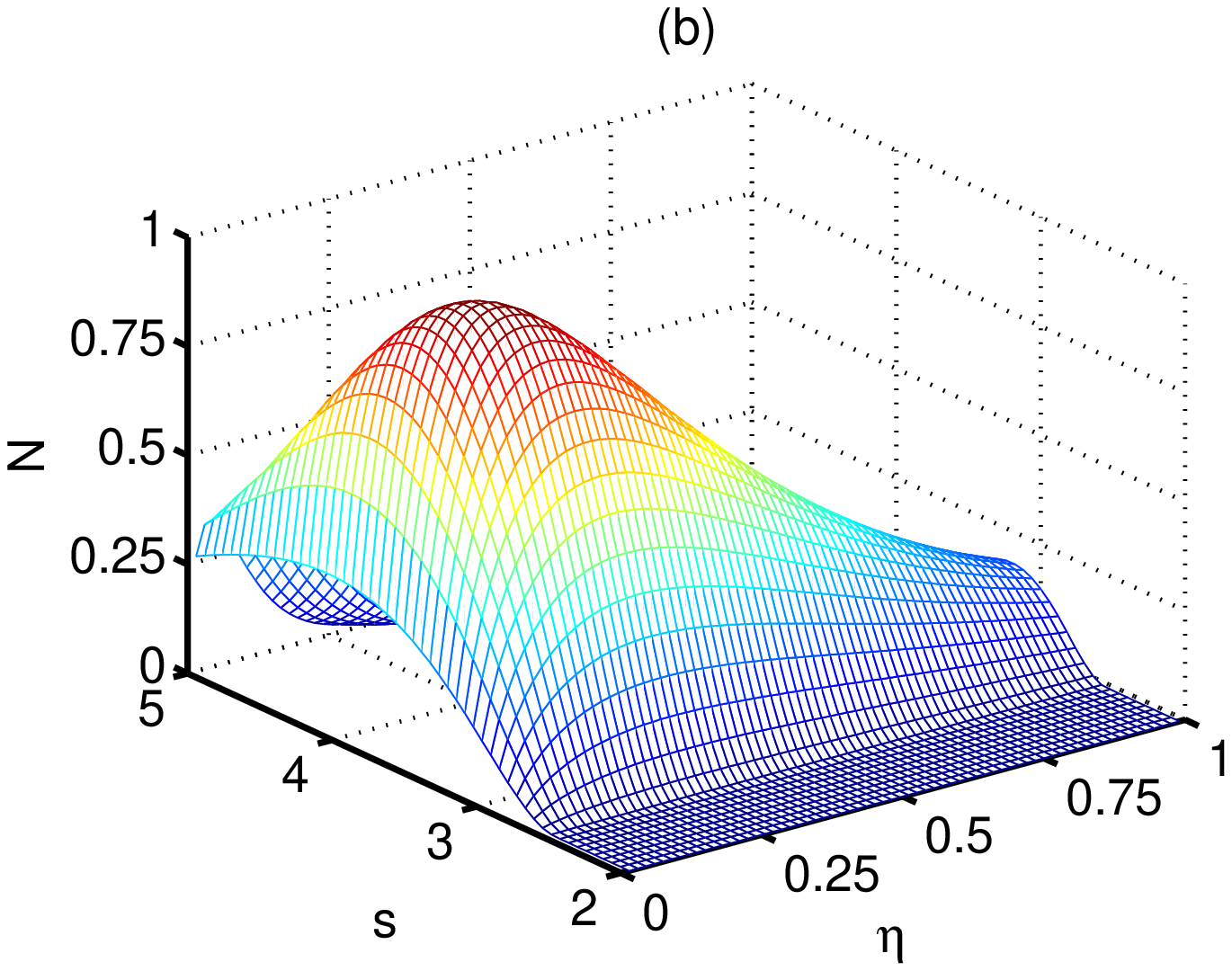}
\caption{(Color online) The non-Markovianity for the dephasing model. Panel (a) is the non-Markovianity vs. the parameters $t$ and $s$, the coupling constant $\eta=0.6$. Panel (b) is the non-Markovianity vs. the parameters $s$ and $\eta$ when the evolution time $t=3$.  The non-Markovianity measure is normalized to unity. }\label{ershenspin}
\end{figure}

\section*{Appendix B: The non-Markovianity of the dephasing model}\label{appendixb}

The memory effects (non-Markovianity) for the dephasing model associated with the quantum channel capacity $Q(\Phi)$ can be defined by \cite{Haikka,nonmar}
\begin{equation}
N=\int_{\partial_t{Q(\Phi )}>0}dt\partial_t{Q(\Phi )},\label{spinnonmarkovian}
\end{equation}
where quantum channel capacity $ Q(t)=1-H_{2}( \frac{1+e^{-\gamma (t)}}{2})$
with $H_{2}(\cdot )$ being the binary Shannon entropy. Due to Eq. (\ref{spinnonmarkovian}), the non-Markovian regime is determined by $\partial_t{Q(\Phi )}>0$. It is easy to find that 
$\partial_t{Q(\Phi )}>0$ means $\gamma ^{\prime }(t)<0$, based on which we plot the Markovian and non-Markovian regions in Fig. 3. The non-Markovianity versus the parameter $s$ and $t$ is plotted in Fig. \ref{ershenspin} (a). Comparing with Fig. \ref{spintu1}, one can find that the blank regime in the Fig. \ref{spintu1} corresponds to the non-Markovian regime.
In Fig. \ref{ershenspin} (b), we plot the non-Markovianity versus the Ohmic parameter $s$ and coupling constant $\eta$ when the evolution time $t=3$. Comparing with the Fig. \ref{spintu2}, one can easily find that the quantum evolution is accelerated within the non-Markovian regime based on the ML-QSL for $\beta=1/\sqrt{2}$ (the upper layer in Fig. \ref{spintu2}). However, for the lower layer in Fig. \ref{spintu2} ($2\beta\sqrt{1-\beta^2}=0.5$), the quantum evolution can also be accelerated in the Markovian regime. This implies the initial-state dependence.

\end{document}